\documentclass[a4paper,11pt]{article}
\usepackage{pos}
\usepackage{slashed,nicefrac,xspace}

\newcommand{\half}{\nicefrac{1}{2}\xspace}

\title{Revisiting experimental mass limits on HECOs using Dyson-Schwinger resummation}
\ShortTitle{HECO production with Dyson-Schwinger resummation}

\author[a]{Jean Alexandre}
\author[a,b]{Nick E. Mavromatos}
\author[c]{Vasiliki A. Mitsou}
\author*[c]{Emanuela Musumeci}
\affiliation[a]{Theoretical Particle Physics and Cosmology group,
Department of Physics, King’s College London,\\
London WC2R 2LS, UK}

\affiliation[b]{Physics Division, School of Applied Mathematical and Physical Sciences,
National Technical University of Athens, 15780 Zografou Campus, Athens, Greece}
\affiliation[c]{
Instituto de Física Corpuscular (IFIC), CSIC – Universitat de València,\\
C/ Catedrático José Beltrán 2, 46980 Paterna (Valencia), Spain}

\emailAdd{jean.alexandre@kcl.ac.uk}
\emailAdd{nikolaos.mavromatos@cern.ch}
\emailAdd{vasiliki.mitsou@ific.uv.es}
\emailAdd{emanuela.musumeci@ific.uv.es}

\abstract{High-Electric-Charge compact Objects (HECOs) appear in several theoretical particle physics models beyond the Standard Model, and are actively searched for in current colliders, such as the Large Hadron Collider (LHC). In such searches, mass bounds of these objects have been placed, using Drell-Yan and Photon-Fusion processes at tree level so far. However, such estimates are not reliable, given that, as a result of the large values of the electric charge of the HECO, perturbative QED calculations break down. We present a Dyson-Schwinger-like resummation scheme, which allows for a large gauge coupling and thus makes the computation of the pertinent HECO-production cross sections reliable, thus allowing us to extract improved mass bounds for such objects from ATLAS and MoEDAL searches.}

\FullConference{42nd International Conference on High Energy Physics (ICHEP2024)\\
18-24 July 2024\\
Prague, Czech Republic\\}

\begin{document}
\maketitle

\section{Introduction}

The existence of High Electric Charge Objects (HECOs) appear in numerous theoretical scenarios beyond the Standard Model, such as Q-balls~\cite{coleman,kusenko}, up-down-quark aggregates~\cite{udaggr}, strange-quark matter~\cite{smatter} and black hole remnants in extra-dimensional models~\cite{bhrem,cbhrem}. Such objects can be produced at colliders via Drell-Yan (DY) and Photon-Fusion (PF)~\cite{Song:2021vpo,Baines:2018ltl}. Both ATLAS and MoEDAL~\cite{MoEDAL:2009jwa,dcmp} collaborations at the LHC are actively searching for these objects~\cite{Altakach:2022hgn}. While current searches rely on tree-level predictions, the large electric coupling, $g=ne$, causes the failure of perturbation theory, requiring resummation, or in general non-perturbative lattice treatments.

In this study, HECOs are modeled as spin-\half Dirac fermions or scalars with electric charges $n \gtrsim 11$. Using Dyson-Schwinger (DS)-like techniques~\cite{Alexandre:2023qjo,Alexandre:2024}, previously developed for magnetic monopoles~\cite{AM}, one can get an effective Lagrangian for both spin-0 and spin-\half HECOs, which leads to reliable mass bounds for these objects, extracted from current collider searches. Here we restrict ourselves to LHC searches  carried out by ATLAS and MoEDAL experiments. An important point for the validity of this approach is the existence of non-trivial Ultra-Violet (UV) fixed points of the Renormalization Group (RG) for these (resummed) field theories, as we discuss next. \color{black}

\section{Scalar HECOs}
\subsection{Effective Lagrangian and Quantum Corrections}
We assume that scalar HECOs only interact with photons in Drell-Yan processes, following~\cite{Song:2021vpo}.
We start with the bare Lagrangian that describes a scalar field interacting with a gauge field~\cite{Alexandre:2024}\color{black}:
\begin{equation}
\mathcal{L}_{\text{bare}} = \frac{1}{2} A^\mu \left[\eta_{\mu\nu}\Box - (1-\lambda)\partial_\mu \partial_\nu\right] A^\nu + D_\mu \phi (D^\mu \phi)^\ast - m^2 \phi \phi^\ast - \frac{h}{4} (\phi \phi^\ast)^2,
\end{equation}
where $D_\mu = \partial_\mu + igA_\mu$ and $\lambda$ is a gauge-fixing parameter. The reader should notice the presence of scalar-HECO non-trivial self interactions, with (bare) coupling $h$. This is crucial for the existence of a non-trivial UV fixed point of the pertinent RG, as discussed in detail in \cite{Alexandre:2024}, and reviewed briefly below. The dressed Lagrangian, accounting for quantum corrections, becomes
\begin{equation}
\mathcal{L}_{\text{eff}} = \frac{1}{2} A^\mu \left[\eta_{\mu\nu}\Box - (1-\lambda)\partial_\mu \partial_\nu\right] A^\nu + \frac{\omega}{2} A^\mu \left[\eta_{\mu\nu}\Box - \partial_\mu \partial_\nu\right] A^\nu 
+ \mathcal{Z}^2 D_\mu \phi (D^\mu \phi)^\ast - M^2 \phi \phi^\ast - \frac{H}{4} (\phi \phi^\ast)^2,
\end{equation}
where $\omega$ and $\mathcal Z$ denote the wavefunction renormalization of the photon and scalar HECO fields, respectively, while $M$ and $H$ denote the renormalzed mass and HECO-self-interaction coupling, respectively.
Renormalized parameters and photon and scalar-HECO wavefunction renormalization depend on the transmutation mass running scale $k$, in the dimensional-regularization approach of \cite{Alexandre:2024}. For simplicity, the gauge coupling $g$ is assumed independent of $k$. The rescaling of the fields $A_\mu \to A_\mu/\sqrt{1+\omega}$ and $\phi \to \phi/\mathcal{Z}$ leads to the canonically normalized Lagrangian:
\begin{equation}
\tilde{\mathcal{L}}_{\text{eff}} = \frac{1}{2} A^\mu \left[\eta_{\mu\nu}\Box - \frac{1-\lambda + \omega}{1+\omega} \partial_\mu \partial_\nu \right] A^\nu + (\partial_\mu \phi + i \tilde{g} A_\mu \phi)(\partial^\mu \phi^\ast - i \tilde{g} A^\mu \phi^\ast)
- \tilde{M}^2 \phi \phi^\ast - \frac{\tilde{H}}{4} (\phi \phi^\ast)^2,
\end{equation}
where the effective parameters are given by:
\begin{equation}
\tilde{g} = \frac{g}{\sqrt{1+\omega}}, \quad \tilde{H} = \frac{H}{\mathcal{Z}^4}, \quad \tilde{M}^2 = \frac{M^2}{\mathcal{Z}^2}.
\end{equation}

\subsection{Resummation of Quantum Corrections}
The DS-like resummation scheme is based on the dressed propagators and vertices in a one-loop approximation:
\begin{equation}
G(p) = \frac{i}{\mathcal{Z}^2 p^2 - M^2}, \quad D_{\mu\nu}(q) = \frac{-i}{(1+\omega)q^2} \left( \eta_{\mu\nu} + \frac{1+\omega - \lambda}{\lambda} \frac{q_\mu q_\nu}{q^2} \right).
\end{equation}
Quantum corrections, such as $M^2 - m^2$, $\mathcal{Z} - 1$ and $\omega$, are assumed to be momentum-independent, but depend on the aforementioned scale $k$ introduced by dimensional regularization. The set of self-consistent equations for the dressed quantities in $d=4-\epsilon$ dimensions is:
\begin{align}
&1 - \frac{m^2}{M^2} = \frac{g^2}{8\pi^2 \mathcal{Z}^2} \left( 1 + \frac{H}{g^2 \mathcal{Z}^2} \right) \ln \left( \frac{k \mathcal{Z}}{M} \right), \qquad
\mathcal{Z}^2 = 1 + \frac{g^2}{8\pi^2} \frac{2-\omega}{1+\omega} \ln \left( \frac{k \mathcal{Z}}{M} \right), \nonumber \\
&\omega = \frac{g^2}{24\pi^2 \mathcal{Z}^2} \ln \left( \frac{k \mathcal{Z}}{M} \right), \qquad
H = h + \frac{3}{8\pi^2} \left( -\frac{H^2}{\mathcal{Z}^4} + g^4 (2\mathcal{Z}^2 - 1) + \frac{Hg^2}{\mathcal{Z}^2} \right) \ln \left( \frac{k \mathcal{Z}}{M} \right).
\end{align}

\vspace{0.2cm}

\subsection{Fixed Point and HECO Mass}
We seek a fixed point solution $(\omega^\star, \mathcal{Z}^\star, H^\star)$ for large  $k \to \Lambda$,  approaching the UV cutoff $\Lambda$,  which defines the energy scale below which the effective field theory is valid. The study in \cite{Alexandre:2024}, which generalised appropriarely the DS-like resummation for spin-\half Dirac HECOs of \cite{Alexandre:2023qjo} to the spin-0 HECO case, 
leads to a consistent set of UV fixed-point parameters (denoted by a $\star$):\color{black}
\begin{align}
(\mathcal{Z}^\star)^2 &= \frac{1+\omega^\star}{1 - 5\omega^\star + 3 (\omega^\star)^2}, \,\,\,\,\,\,
H^\star = h + 9\omega^\star \left( -\frac{(H^\star)^2}{g^2 (\mathcal{Z}^\star)^2} + g^2 (\mathcal{Z}^\star)^2 (2(\mathcal{Z}^\star)^2 - 1) + H^\star \right).
\end{align}
The range for $\omega^\star$ that satisfies the constraints is $0 \leq \omega^\star \lesssim 0.11$ and the mass of the scalar HECO in this limit is:
\begin{equation} \label{mass}
\tilde{M} \simeq \Lambda \exp \left( - \frac{32\pi^2}{h} \right),
\end{equation}
indicating that for a consistent fixed point, strong self-interactions among scalar HECOs are required, assuming an UV cutoff $\Lambda$. As discussed in detail in \cite{Alexandre:2024}, 
self consistency for the existence of the fixed point solution requires 
strong HECO-self-interaction couplings, specifically
$ h \gtrsim 12.12g^2$. In our analysis here we shall assume saturation of this inequality.

\subsection{Interaction of Scalar HECOs and Experimental Assumptions}
Given the form of the photon propagator for $\lambda=1$, the fixed point we describe is analogous to an effective gauge choice, $\lambda_{\text{eff}} = \frac{1}{1 + \omega^\star}$, which 
leads to a gauge-fixed (non-gauge invariant) effective Lagrangian. As discussed in detail in \cite{Alexandre:2023qjo,Alexandre:2024}, this effective gauge-non invariance of the DS-like resummed theory plays a crucial in avoiding the usual equality constraint between the HECO-photon-vertex and scalar-HECO wavefunction renormalizations, which in gauge invariant Quantum Electrodynamics (QED) theories would stem from the validity of the relevant  Ward identities. The charge correction for the HECO, $g \to g\mathcal{Z}^\star$, does not cancel due to the resummation effects, in contrast to standard QED. Consequently, we define the charge as:
\[
g_{\text{HECO}} = g \mathcal{Z}^\star.
\]
Following earlier works, the HECO electromagnetic coupling can be written as $g = n e$, where $n \in \mathbb{Z}$ and $|n| \gtrsim 11$~\cite{Alexandre:2023qjo}. 
The Feynman rules at the UV fixed point $\omega^\star$ for scalar HECOs are:
\begin{itemize}
    \item Scalar-photon vertex: $-ig \mathcal{Z}^\star \simeq -ig \sqrt{1 + \frac{8g^2}{h}}$
    \item Scalar self-interaction vertex: $-i\frac{H^\star}{{\mathcal Z}^{\star 4}} \simeq -i\frac{h}{4}\left(1 + \frac{8g^2}{h}\right)^{-2}$
       \item Gauge boson propagator:$ \frac{-i}{p^2 - i\epsilon} \left(\eta_{\mu\nu} + \frac{4g^2}{3h}\, \frac{p_\mu p_\nu}{p^2}\right)\,, \quad \epsilon \to 0^+$ 
    
    \item Charged scalar propagator: $ \frac{i}{p^2 - \tilde{M}^2 + i\epsilon}, \quad \epsilon \to 0^+$   
\end{itemize}
Here, $\tilde{M}$ is the scalar HECO mass, defined by Eq.~(\ref{mass}), with $h\simeq 12.12g^2$.

\section{Spin-\half HECOs}
A detailed treatment of fermion HECO interactions and DS-like resummation can be found in~\cite{Alexandre:2023qjo, Musumeci:2024prf}. Both DY and PF production mechanisms are considered, with the $Z^0$ boson interaction included in DY for fermions. The HECO-photon interaction is governed by the following Feynman rules:
\begin{itemize}
\item Running mass: ${\tilde M}(\Lambda) = \Lambda \exp\left( -\frac{2\pi}{\alpha^\star} (Z^\star - 1) \right)$,
\item HECO fermion propagator: $G^{\rm eff} = i \frac{\slashed p + \tilde{M}(\Lambda)}{p^2 - \tilde{M}(\Lambda)^2}$,
\item $\gamma$ propagator: $\Delta^{\rm eff}_{\mu\nu} = \frac{-i}{q^2}\left(\eta_{\mu\nu}+ \omega^\star\, \frac{q_\mu q_\nu}{q^2}\right)$
\item Photon-HECO vertex: $\Gamma^{\rm eff}_\mu = g \, Z^\star \gamma_\mu$,
\end{itemize}
where $\hat\alpha^\star = \frac{g^2/4\pi}{1+\hat\omega^\star}$, $Z^\star$ is the wavefunction renormalisation, and $\omega^\star = \frac{4}{3}\left(1 - \frac{1}{Z^\star}\right)$.
For additional interactions with the $Z^0$ boson:
\begin{itemize}
\item $g^2 \to \hat{g}^2 \equiv g^2 + 3g'^2/4$, where $g'$ is the $Z^0$-HECO coupling,
\item $\hat{Z}^\star = \frac{2}{9}(3+\eta) \left( 1 + \sqrt{1 - \frac{9\eta}{(3+\eta)^2}} \right)$ with $\eta = g^2/\hat{g}^2 < 1$,
\item $\hat{\omega}^\star = \frac{4}{3} \eta \left( 1 - \frac{1}{\hat{Z}^\star} \right)$.
\end{itemize}

\section{Effects on Cross Section and Mass Constraints}
The resummation effects were implemented in Universal Feynrules Output models~\cite{ufo}, compatible with Monte Carlo event generators like \textsc{MadGraph}~3.5.4~\cite{madgraph}. The model focuses on DY and PF processes and incorporates the Feynman rules derived from the effective Lagrangian at the UV fixed point. The model's cross-section estimates were validated using \textsc{Mathematica}~13.0.1--- specifically the \textsc{Feyncalc} package~\cite{Feyncalc} --- by comparing the results from \textsc{MadGraph}~3.5.4 with the analytical ones. New parameters, including charge multiplicity $n$ and the cutoff energy scale $\Lambda$, are introduced into the UFO models for both spin-0 and spin-\half HECOs. These supplement standard inputs like center-of-mass energy and Parton Distribution Functions (PDFs). 

\begin{figure}
   \includegraphics[width=0.5\linewidth]{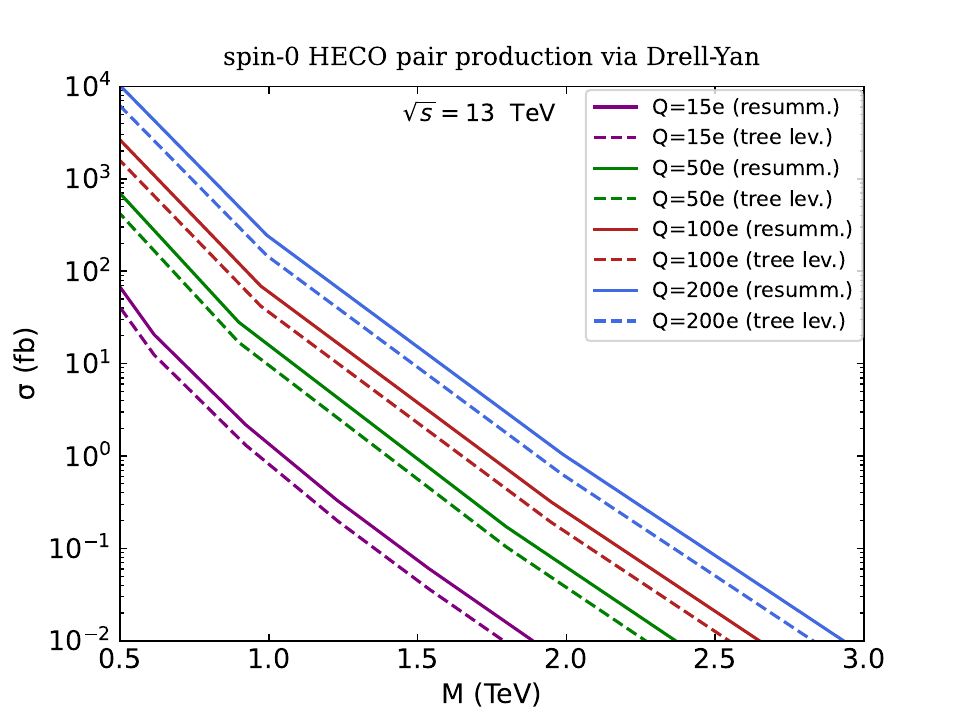}
    \includegraphics[width=0.5\linewidth]{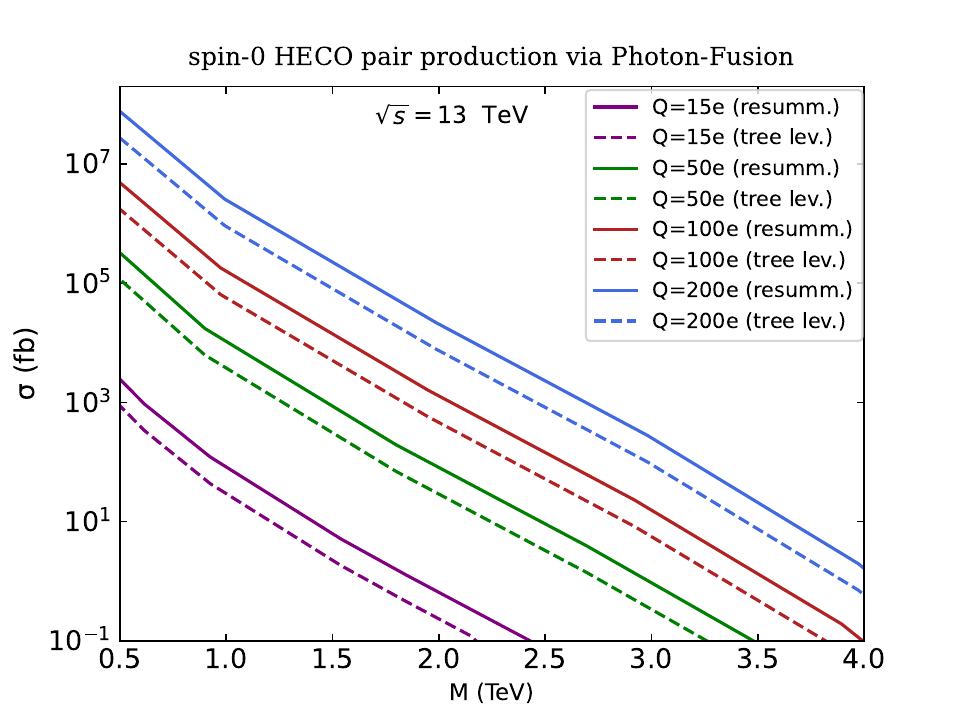}
    \includegraphics[width=0.5\linewidth]{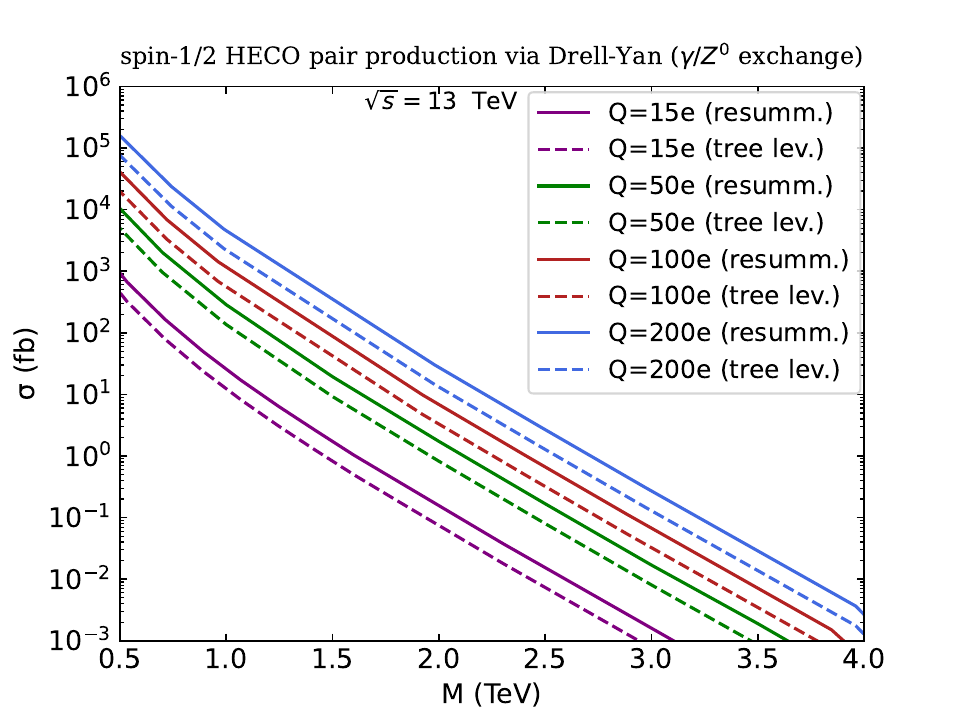}
    \includegraphics[width=0.5\linewidth]{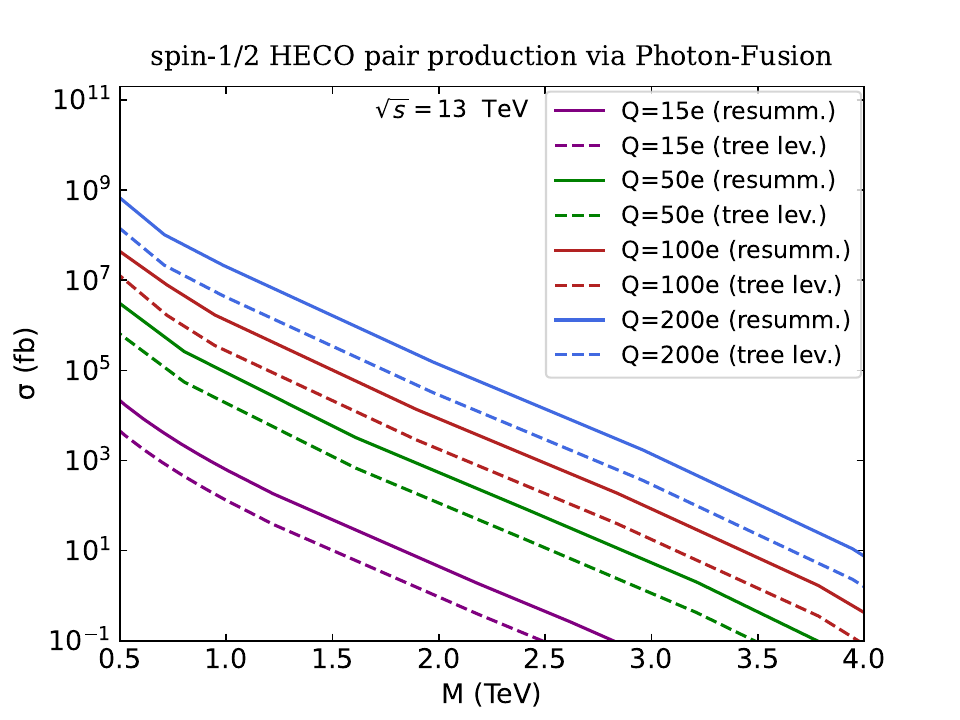}
    \caption{\footnotesize Comparison of cross sections before (dashed line) and after (solid line) resummation as a function of the HECO mass $M$ for electric charges $Q = 15e$, $50e$, $100e$, and $200e$, for both spin-0 (top) and spin-\half (bottom) HECOs.}
    \label{fig:aftervsbefore}
\end{figure}

Resummation effects show a notable impact on production cross sections  as shown in Fig.~\ref{fig:aftervsbefore}., with calculations performed for $pp$ collisions at $\sqrt{s}=13~$TeV. The \texttt{NNPDF23} PDF is used for DY, while \texttt{LUXqed17} is employed for PF.
In the scalar case, resummation enhances production by factors of $\sim 1.66$ for DY and $\sim 2.76$ for PF. For the fermion case, the factors are $\sim 2.1 $ for DY and $\sim 4.75$ for PF. Notably, the cross-section values for the PF process exceed those of DY at LHC energies. This difference arises from their scaling behaviors, with cross-section dependence proportional to $ n^2 $ for DY and $ n^4 $ for PF.
The study further includes the re-interpretation of experimental data from the ATLAS~\cite{ATLAS:2015tyu,ATLAS:2023esy} and MoEDAL~\cite{MoEDAL:2021mpi,MoEDAL:2023ost} collaborations. By applying DS resummation techniques, we derive more stringent mass limits for spin-\half HECOs compared to those obtained from tree-level calculations. In~\cite{Alexandre:2023qjo}, the 95\% confidence level (CL) mass limits for various spin-\half HECOs are presented, consistently demonstrating enhanced constraints. This increase in mass limits spans a broad range of values, reaching up to 30\%.The limits for spin-0 HECOs are still forthcoming, but their behavior is expected to be similar~\cite{Alexandre:2024}.

\section*{Acknowledgments}
E.M.\ acknowledges financial support from the MICIN with funding from the European Union NextGenerationEU and from Generalitat Valenciana (GV) via the Programa de Planes Complementarios de I+D+i (PRTR 2022) Project \textit{Si4HiggsFactories}, ASFAE/2022/015. The work of J.A. and N.E.M. is supported in part by the UK Science and Technology Facilities research Council (STFC) under the research grant ST/X000753/1. \color{black} V.A.M.\ and E.M.\ acknowledge support by GV via the Excellence Grant Prometeo CIPROM/2021/073 and by the Spanish MICIN / AEI / 10.13039/501100011033 and the European Union / FEDER via  grant PID2021-122134NB-C21.

\bibliographystyle{JHEP-vaso-notitle.bst}
\setlength{\bibsep}{2pt}
\small
\bibliography{ichep_heco}

\end{document}